\documentclass[12pt]{JHEP3}
\usepackage{amsmath,epsf,amssymb,latexsym,cite,graphics}
\usepackage[matrix,arrow,frame,import,curve,color]{xy}

\newif\iffigs\figstrue

\DeclareFontFamily{U}{rsf}{}
\DeclareFontShape{U}{rsf}{m}{n}{
  <5> <6> rsfs5 <7> <8> <9> rsfs7 <10-> rsfs10}{}
\DeclareMathAlphabet\Scr{U}{rsf}{m}{n}

\def\pplogo{\vbox{\kern-\headheight\kern-29pt
\halign{##&##\hfil\cr&{
\ppnumber}\cr\rule{0pt}{2.5ex}&\ppdate\cr}
}}
\makeatletter
\def\ps@firstpage{\ps@empty \def\@oddhead{\hss\pplogo}%
  \let\@evenhead\@oddhead 
}
\def\maketitle{\par
 \begingroup
 \def\thefootnote{\fnsymbol{footnote}}
 \def\@makefnmark{\hbox{$^{\@thefnmark}$\hss}}
 \if@twocolumn
 \twocolumn[\@maketitle]
 \else \newpage
 \global\@topnum\z@ \@maketitle \fi\thispagestyle{firstpage}\@thanks
 \endgroup
 \setcounter{footnote}{0}
 \let\maketitle\relax
 \let\@maketitle\relax
 \gdef\@thanks{}\gdef\@author{}\gdef\@title{}\let\thanks\relax}
\makeatother

\def\O{\Scr{O}}
\def\cH{\Scr{H}}
\def\Z{\Scr{H}}

\def\C{{\mathbb C}}

\def\P{{\mathbb P}}
\def\R{{\mathbb R}}
\def\Z{{\mathbb Z}}

\def\GU{\operatorname{U{}}}

\def\Hess{\operatorname{Hess}}

\def\p{\partial}

\def\cN{{\Scr N}}

\def\ff#1#2{{\textstyle\frac{#1}{#2}}}

\def\ep{\epsilon}

\def\la{\langle}
\def\ra{\rangle}

\def\sigmab{\bar{\sigma}}
\def\Wb{\overline{W}}

\def\rhot{\widetilde{\rho}}
\def\omegat{\widetilde{\omega}}
\def\gammat{\widetilde{\gamma}}
\def\sigmah{{\hat{\sigma}}}

\title{Non-Local Observables in the $A$-Model}
\author {Ilarion V. Melnikov \\
\normalsize Enrico Fermi Institute \\
\normalsize University of Chicago \\
\normalsize Chicago, IL 60637, USA\\
Email:  \email{lmel@theory.uchicago.edu}
}
\abstract{We compute correlators of non-local observables in a large class of $A$-twisted massive Landau-Ginzburg and 
gauged linear sigma models by localization to the discrete vacua.  As an application, we present two topological field theories with
identical chiral rings and correlators of local observables, which nevertheless differ in the correlators of non-local observables. }

\preprint{EFI-07-03 \\ hep-th/0701186}
\keywords{Topological Field Theories, Sigma Models}

\begin{document}


\section{Introduction}
The computation of correlation functions of local operators pays the bills of many 
a practicing quantum field theorist.  These correlators  contain a wealth of information 
about a quantum field theory, and  there are well-developed techniques for a 
proper regularization and  renormalization of these objects.  Of course, in these theories
it is possible to write down non-local operators as well.  Perhaps the most familiar class of 
such operators is given by Wilson lines in a gauge theory.  Correlators of such operators are 
more difficult to compute, but their computation carries substantial rewards, especially on 
topologically non-trivial space-times, where they are often sensitive to topological properties of
the underlying space-time that would be difficult or impossible to discern from local  observables 
alone.

Topological quantum field theories are richly endowed with  non-local
observables.  Whether it is Chern-Simons gauge theory on a three-manifold \cite{W:qftjp},
Donaldson theory \cite{W:tqft}, or the topological twist of $\cN=4$ Super Yang-Mills theory that
appears in the recent work of Kapustin and Witten on the geometric Langlands program and 
electric-magnetic duality\cite{KW}, answers to intricate geometric (and even number theoretic!) 
questions are encoded in correlators of non-local observables.

This note is devoted to the study of non-local observables in a simple class of two-dimensional topological quantum
field theories:  twisted massive Landau-Ginzburg theories and topological sigma 
models with compact toric target spaces.   We will show that in these theories correlators with insertions of 
one-form non-local observables are readily computable by simple localization techniques and yield additional 
information about the quantum field theory. The geometric significance of these new correlators is, as yet, unclear, and we believe that
for a proper geometric interpretation we will need to generalize the localization techniques to the topological
field theory coupled to topological gravity.  Nevertheless, we believe that our results are of interest as 
an \'etude in exactly soluble field theory, as a study of some new properties of the topological sigma 
model, and as a reconnaissance in the direction of the more interesting case of coupling
these ``massive'' topological field theories to two-dimensional gravity.

We end this section with a brief outline of the rest of the note.  We will begin with a brief review of
general properties of cohomological topological quantum field theory and topological observables, and 
we will illustrate them in the case of a simple example:  the twisted massive Landau-Ginzburg model.   
Next, in section \ref{s:lgnonloc} we will present one of our main results:  the computation of correlators in
the twisted massive Landau-Ginzburg theory with insertions of one-form non-local operators.  In section 
\ref{s:glsm} we will review the relation---via the gauged linear sigma model---between the topological 
sigma model with a compact toric target-space and a particular massive Landau-Ginzburg theory.  This 
will enable us to adapt the results of section \ref{s:lgnonloc} to compute new correlators in these topological 
sigma models.  We will apply our general formulas to two examples in section \ref{s:exmpl}, and demonstrate 
one use of the non-local operator insertions:  they can distinguish models that may otherwise seem equivalent.  
We will wrap up in section \ref{s:discus} with a discussion of some general properties of the new correlators.
The Appendix explores properties of the two-form observables in the Landau-Ginzburg theory. 

\section{A Review of Cohomological Topological Field Theories} \label{s:cohorev}
There are a number of excellent reviews of this beautiful subject \cite{BT:localization,CMR:ym,W:coho,DVV:tftnotes}, 
and we will not try to cover the details of Cohomological Topological Field Theory (CTFT) in any detail.
Instead, we hope to provide the reader with a sufficient reminder to place our work in its proper context.

Typically, a field theory on some fixed curved space-time contains detailed information
about the geometry of the space.  After all, classical particles follow geodesics, and field equations 
and depend sensitively on the metric.  This dependence is encoded by the energy-momentum tensor of the
theory.  By definition, a Topological Field Theory (TFT) is not sensitive to small changes in the space-time 
metric and only involves coarser properties of the space-time.  One way to obtain a TFT is to 
pick an action that does not involve the spacetime metric.\footnote{This is not just a matter of defining a classical 
action that is metric  independent.  One must also demonstrate that the regularization procedure one uses to 
render the QFT sensible does not  re-introduce metric dependence.} Chern-Simons theory is a prime example of this sort 
of theory.  The Cohomological approach is different.  In this case the action may depend on the metric, 
but the theory possesses a BRST-like symmetry which renders this dependence trivial.  

Many CTFTs can be constructed by the elegant procedure of ``twisting'' \cite{W:topsig,W:mirtop,W:tqft}:  one begins 
with a field theory with extended supersymmetry and  modifies the coupling of the fermions to gravity so that at least
one of the supercharges becomes a space-time scalar operator.  This operator squares to zero, and its cohomology
defines the set of observables.  The theories we will study below are of this sort.

\subsection{Action and Local Observables}
For our purposes a (Lagrangian) CTFT on a manifold $M$  with a Riemannian metric $g$  is specified by:
a set of fields $\phi$ with a local action $S[\phi,g]$; a measure for the path integral $D[\phi]$, and a space-time scalar 
anti-commuting  operator $Q$ generating transformations $\delta \phi = \{Q, \phi\}$ \footnote{$\{Q,\phi\}$  is a short-hand for
$Q\phi \mp \phi Q$, with the sign depending on whether $\phi$ is bosonic ($-$) or fermionic ($+$).}  such that
\begin{eqnarray}
\label{eq:Qprop}
\{Q, S\} & = & 0, \nonumber\\ 
 T_{ab} & = & \{Q,\cdot\}, \nonumber\\
\int D[\phi] \{Q, \cdot\} & = & 0,
\end{eqnarray}
where $T_{ab}$ is the energy-momentum tensor: $T_{ab} = - \delta S /\delta g^{ab}$.
In most CTFTs the ``$Q$-exactness'' of $T_{ab}$ follows from a  particular form of the action:
\begin{equation}
S[\phi,g] = S_{\text{top}} [\phi] + \{Q, I[\phi,g] \},
\end{equation}
where $S_{\text{top}}[\phi]$ is a purely topological term, while $I$ contains the dependence on the chosen
metric.  The theories we will study below have this form of the action.

While the full theory will depend on details of the chosen metric $g$, we can obtain a consistent topological
sector of the theory by restricting computations to correlators of $Q$-closed operators, i.e. operators satisfying
$\{Q,\O\} =0$. We will refer to these as {\em observables}.  The properties of the CTFT given in eqn. (\ref{eq:Qprop}) 
ensure that correlators of observables are independent of the metric $g$ and only depend on the $Q$-cohomology 
classes of the observables.  

The simplest class of observables is obtained by restricting to local $Q$-closed operators.  Since the energy-momentum
tensor of the CTFT is $Q$-exact, the correlators of local observables, $\la \O_1(x_1) \cdots \O_k(x_k)\ra$ are independent 
of the positions $x_i$, implying, in particular, that any singular terms in the OPE  $\lim_{x\to 0} \O_1(x) \O_2(0)$ are 
$Q$-trivial.  This allows a choice of zero contact terms in the projected theory, and the OPE gives the set of local observables 
a ring structure.  In the models we will consider this will be a finite ring, and, by analogy with $N=(2,2)$ SUSY SCFTs, we will 
refer to it as the {\em chiral ring} of the CTFT. 

\subsection{Non-Local Observables via { Descent}}
 The local observables do not exhaust the set of topological observables, and there is an elegant procedure going back to
the original work of Witten \cite{W:tqft} that produces non-local topological observables from local ones.  This 
procedure, which we will now describe, has come to be known as {\em descent}.
 
Let $\O$ be a local observable in a CTFT defined on a manifold $M$.  Since translations are generated by the $Q$-exact energy-momentum 
tensor, it is clear that
\[ d \O = \{Q, \O_{(1)}\} \]
for some one-form valued operator $\O_{(1)}$.   Given a $1$-cycle $C \in M$ the non-local operator
$\int_C \O_{(1)}$ is $Q$-closed and thus an observable. This descent procedure can be iterated:
given a $k$-form valued operator $\O_k$,
\[ d\O_k = \{Q, \O_{(k+1)}\} \]
for some $k+1$-valued operator $\O_{(k+1)}$, and  for any $k+1$-cycle $C_{k+1}$, $\int_{C_{k+1}} \O_{(k+1)}$ is an 
observable.  When we need to distinguish between the non-local observables, we will refer to observables of the 
form $\int_{C_k} \O_k$ as {\em $k$-form observables.}

The observables obtained by descent have three important properties:
\begin{itemize}
\item[-] by Stokes' theorem, the $Q$-cohomology class of $\int_{C_{k+1}} \O_{(k+1)}$ only depends on the 
homology class of $C_{k+1}$.
\item[-]
Descendants of a $Q$-trivial operator $\O_k$ are $Q$-trivial.  
Indeed, if $\O_{k} = \{Q, V\}$ for some $V$, then, since $d$ and $Q$ commute, we have
\[  \O_{(k+1)} = d V + \{Q,U\} \]
for some operator $U$, and, as expected, $\int_{C_{k+1}} \O_{k+1}$ is $Q$-exact.
\item[-]  The operators $\O_{\dim(M)}$ obtained by descent may be used to deform the action 
$S \to S + \lambda \int_M \O_{\dim(M)}$ while keeping $S$ local and $Q$-closed. 
\end{itemize}

So far, we have discussed CTFTs and their observables in a very general fashion.  In what follows, we will see all of
these concepts illustrated in a set of concrete and fairly simple examples. 

\subsection{Topological Landau-Ginzburg Models} 
We will now study what is perhaps the simplest CTFT: a massive twisted Landau-Ginzburg (L-G) model defined 
on a Riemann surface $\Sigma_h$ of genus $h$.  These theories were first considered by Vafa \cite{Vafa:tlg}, and our 
introduction to these models will follow his original presentation.  

These models are constructed by twisting the $N=(2,2)$ SUSY L-G models, and it should come as no surprise that
that the field content of such a model is organized into multiplets $\Phi_a$, with a structure familiar from the $N=(2,2)$ theory.
Each multiplet $\Phi_a$ contains
\begin{itemize}
\item $\sigma_a$:  a complex bosonic scalar;
\item $\theta_a$, $\chi_a$: fermionic scalars;
\item $\rho_a$: a fermionic one-form.
\end{itemize}

The action for the theory with $r$ multiplets depends upon the superpotential $W(\sigma)$, a holomorphic 
function of the bosonic scalar fields:
\begin{eqnarray}
\label{eq:LGaction}
S & = & \int_{\Sigma_h} \left\{
\sum_{a=1}^r \left[ d\sigma_a \wedge\ast d \sigmab_a + 2\rho_a\wedge \ast d\theta_a  + 2 i\rho_a\wedge d\chi_a \right]\right. \nonumber\\
~& &~~~~~~\left. +\sum_{a,b=1}^r \left[\ast (|W_{,a}(\sigma)|^2 + 2\chi_a \Wb_{,ab} \theta_b) - i\rho_a\wedge\rho_b W_{,ab}\right]\right\}.
\end{eqnarray}
This theory admits the action of a fermionic scalar $Q$:
\begin{eqnarray}
\label{eq:Qm}
\{Q,\sigma_a \}  &=&  0,  \nonumber\\
\{Q,\sigmab_a\} &=& 2\theta_a, \nonumber\\
\{Q,\theta_a\} &=& 0, \nonumber\\
\{Q,\chi_a\} &=& -W_{,a}(\sigma), \nonumber\\ 
\{Q,\rho_a\}  &=&  -d\sigma_a. 
\end{eqnarray} 
It is easy to show that $Q^2 =0$, $\{Q,S\}=0$, and the action may be written as a sum of 
$S_{\text{top}}$ and $S_{\text{triv}} = \{Q,I\}$, with
\begin{eqnarray}
\label{eq:LGactionsplit}
S_{\text{top}} & = & i \int_{\Sigma_{h}} \left( 2\rho_a \wedge d\chi_a - \rho_a\wedge\rho_b W_{,ab}\right),\nonumber\\
I & = & \int_{\Sigma_{h}} \left( -\ast \chi_a \Wb_{,a} -\rho_a \wedge\ast d\sigmab_a \right).
\end{eqnarray}
The equations of motion which follow from $S$ are
\begin{eqnarray}
\label{eq:eom}
d\chi_a & = & W_{,ab} \rho_b -i \ast d\sigma_a, \nonumber\\
d\rho_a & =& \{Q, -\ff{i}{2} \ast \Wb_{,a}\} = -i\ast \Wb_{,ab} \theta_b, \nonumber\\
d\ast\rho_a & = & - \ast \Wb_{,ab} \chi_b, \nonumber\\
d\ast d\sigmab_a & =& -\ast W_{,ab} \Wb_{,b} - i \rho_b \wedge \rho_c W_{,bca}, \nonumber\\
d\ast d\sigma_a  & = & \ast W_{,b} \Wb_{,ab} -2 \ast \chi_b \theta_c W_{,bca}.
\end{eqnarray}

\subsubsection{The Free Theory}
To develop facility with localization techniques that we will use throughout this note, we will begin with the simple problem of
computing the partition function for the free theory, i.e.  $W = \ff{1}{2} m^{ab} \sigma_a \sigma_b$.  To define
the path integral, we will expand the fields in the eigenmodes of the Hodge-De Rham Laplacian for some fixed metric
$g$ on $\Sigma_h$:
\begin{equation}
\label{eq:Laplacian}
\Delta_d f_k = \lambda_k^2 f_k,~~ f_k \in \Omega^0(\Sigma_h),~ \int_{\Sigma_h} (\ast f_k) f_l = \delta_{kl},~ \lambda_k \neq 0.
\end{equation}
The fields may be expanded as\footnote{To avoid clutter, we suppressed the multiplet index.}
\begin{eqnarray}
\label{eq:fieldslap}
\sigma  & = & \ff{1}{\sqrt{V_g}} \sigma_0 + \sum_k \sigma_k f_k, \nonumber\\
\chi  & = & \ff{1}{\sqrt{V_g}} \chi_0 + \sum_k \chi_k f_k, \nonumber\\
\theta  & = & \ff{1}{\sqrt{V_g}} \theta_0 + \sum_k \theta_k f_k, \nonumber\\
\rho  & = & \sum_{\alpha=1}^{h} (\rho_0^\alpha \omega_\alpha + \rhot_{0\alpha} \omegat^\alpha) +
                  \sum_k \ff{1}{\lambda_k} (\rho_k d f_k + \rhot_k \ast d f_k),
\end{eqnarray} 
where  $V_g$ is the volume of $\Sigma_h$ in the metric $g$, and 
$\{\omega_1, \omegat^1, \ldots, \omega_h, \omegat^h\}$ is a symplectic basis for $H^1(\Sigma_h, \R)$
satisfying
\begin{equation}
\int_{\Sigma_h} \omega_\alpha \wedge \omegat^\beta = \delta^\beta_\alpha, ~ 
\int_{\Sigma_h} \omega_\alpha \wedge \omega_\beta= 0,~
\int_{\Sigma_h} \omegat^\alpha \wedge \omegat^\beta = 0.
\end{equation}
We can now write a regulated measure for the path integral:
\begin{equation} 
D[\text{fields}]_N = \prod_{a=1}^n D[ \Phi_a]_N,
\end{equation}
where for each multiplet we have
\begin{equation}
\label{eq:freemeasure}
D[\Phi]_N = \frac{d^2\sigma_0}{\pi} \frac{d\chi_0 d\theta_0 }{2} \prod_{\alpha=1}^h \frac{d \rhot_{0\alpha} d \rho_0^\alpha}{2i}
                                  \prod_{k<N} \frac{d^2\sigma_k}{\pi} \frac{ d\chi_k d\theta_k d\rhot_k d\rho_k}{4i}.
\end{equation}
Plugging in the mode expansion into the action, we find that $S$ can be written as a sum of the zero-mode and non-zero mode terms, $S_0$
and $S'$:
\begin{eqnarray}
S_0 & = & m^{ac} \bar{m}^{cb} \sigma_{a,0}\sigmab_{b,0} +2 \chi_{a,0} \bar{m}^{ab} \theta_{b,0} + 2i m^{ab} \sum_{\alpha=1}^h \rhot_{a,0\alpha}\rho_{b,0}^\alpha \nonumber\\  
S'     & = & \sum_k \left\{ (\lambda_k^2 \delta^{ab} + m^{ac}\bar{m}^{cb})\sigma_{a,k}\sigmab_{b,k} + 2 \lambda_k(\rho_{a,k}\theta_{a,k} -i \rhot_{a,k} \chi_{a,k}) \right. \nonumber\\
        & ~ &~~~~~~ \left. +2 \bar{m}^{ab} \chi_{a,k} \theta_{b,k} -2i m^{ab} \rho_{a,k} \rhot_{b,k}\right\}.
\end{eqnarray}
It is easy to see that the contributions from the non-zero modes pair up and cancel, and the non-trivial dependence of the partition function
on $m$ is due to an incomplete cancellation among the contributions from the zero modes.  Performing the trivial determinant computations,
we find that the partition function is given by
\begin{equation} Z_N = \int D[\text{fields}]_N e^{-S} = (\det m)^{h-1}. \end{equation}
In particular, $Z_N$ is $N$-independent and we may safely remove the regulator by taking $N \to \infty$. 

\subsubsection{Arbitrary Superpotential}
The simplest way to compute topological correlators for general $W$ is via localization of the path integral on field configurations
annihilated by $Q$.  Localization is a general property of CTFTs \cite{SZ:superlocal} that is particularly easy to
understand in this simple theory \cite{Vafa:tlg}.  Consider rescaling the metric $g$ on $\Sigma_h$ by a constant factor: 
$g\to \lambda g$.  This is a $Q$-exact change in the action of theory, so that, assuming there are no subtleties in defining
the measure, the topological correlators will be $\lambda$-independent and we may compute them in the $\lambda \to \infty$
limit. Expanding out $\{Q,I[\phi,\lambda g] \}$, it is clear that in this limit the path integral will be supported on configurations 
satisfying 
\begin{equation}
\label{eq:localization}
d \sigma^a = 0,~\text{and}~ \frac{\p W}{\p \sigma_a} = 0,
\end{equation}
rendering the saddlepoint approximation to the path integral obtained by expanding the action to 
quadratic order about classical vacua exact!  Assuming that the solutions to  $\ff{\p W}{\p \sigma_a}=0$ are isolated points
$\sigmah \in \C^n$, to compute the partition function we write $\sigma = \sigmah + \sigma'$, repeat the free field theory 
computation from above with $m^{ab} = \ff{\p^2 W}{\p \sigma_a \p \sigma_b} |_{\sigma = \sigmah}$ 
and sum over the vacua.  This yields
\begin{equation}
\label{eq:genpartition}
Z = \sum_{\sigmah} [\det\Hess W]^{h-1}.
\end{equation}
Repeating the localization argument with insertions of local operators, it is easy to convince oneself that
the correlators of local observables are equally simple:
\begin{equation}
\label{eq:loccorrs}
\la \sigma_{a_1}(x_1) \cdots \sigma_{a_k}(x_k)\ra_h  =  \sum_{\sigmah| dW(\sigmah) = 0} [\det\Hess W]^{h-1}
\sigmah_{a_1} \cdots \sigmah_{a_k} ~.
\end{equation}
As expected on general grounds, these correlators are independent of the $x_i$, a fact we will abuse by abbreviating
these insertions as $\la F(\sigma)\ra$.  The Landau-Ginzburg TFT has a finte chiral ring, $\C[\sigma_1,\ldots,\sigma_r]/(W')$, 
and the above formula serves to determine correlators with arbitrary insertions of local observables.

\section{Non-Local Observables in the Landau-Ginzburg TFT} \label{s:lgnonloc}
We will now use descent to obtain a set of non-local observables in the Landau-Ginzburg theory. We start
with a local observable
\begin{equation}
\O_{f(0)} = f(\sigma)(x),
\end{equation}
and note that $d\O_{f(0)} = f_{,a} d\sigma_a = \{Q,-f_{,a} \rho_a\}$.  Thus, we see that 
\begin{equation}
\O_{f(1)} = -f_{,a} \rho_a
\end{equation}
can be used to make the one-form observable $\int_C \O_{f(1)}$ for any closed curve $C \subset \Sigma_h$.  
Repeating the procedure, we see that
\begin{equation}
d\O_{f(1)} = -f_{,ab} d\sigma_a \wedge \rho_b - f_{,a} d \rho_a.
\end{equation}
At first sight, it is not obvious how to write the right-hand side as $\{Q,\cdot\}$.  To make progress,
we use the equations of motion to rewrite $d \rho_a$  as $\{Q,-\ff{i}{2} \ast \Wb_{,a}\}$ and obtain
\begin{equation}
d \O_{f(1)} = \{Q, \ff{1}{2} (f_{,ab} \rho_a \wedge \rho_b + i f_{,a} \ast \Wb_{,a}) \}.
\end{equation}
Thus, $2i \O_{f(2)} = (i f_{,ab} \rho_a\wedge \rho_b - f_{,a} \ast \Wb_{,a})$ is a two-form whose integral over the Riemann
surface yields another non-local observable.  Using the equations of motion, it is easy to verify that operators obtained by 
descent from a $Q$-exact local operator, such as $\O_{W_{,a}(0)}$, are also $Q$-exact.

The two-form observables are interesting in their own right, and we will study some of their features in the Appendix.  We will
outline how localization may be used to compute correlators with two-form observable insertions, and we will verify that these
two-form insertions correspond to deformations of the Landau-Ginzburg superpotential.  However, our primary interest in this
note will be in the correlators involving one-form observables, and it is to these objects that we now turn.

\subsection*{Correlators of One-Form Observables}
We are now ready to deal with the operators of most interest to us:  the ones based on $\O_{f(1)}$.  We will show that correlators of these
operators are just as easy to compute as correlators of their local ancestors.  It is sufficient to consider the case of $f=\sigma_a$, for which the
non-local observables take the form
\begin{equation}
\label{eq:nonlocob}
\gamma_a[C] = \int_C \rho_a.
\end{equation}
It is convenient to choose a basis for $H_1(\Sigma_h,\Z)$ dual to the basis of $H^1(\Sigma_h,\R)$ used above:  we pick a basis
of one cycles $\{C_\alpha, C^\alpha\}$ such that
\begin{equation}
\int_{C_\alpha} \omega_\beta = \delta^\alpha_\beta,~~\int_{C^\alpha} \omegat^\beta = \delta^\beta_\alpha,~~\int_{C_\alpha} \omegat^\beta = \int_{C^\alpha} \omega_\beta = 0,
\end{equation}
and we decompose the non-local observables into $\gamma^\alpha_a = \gamma_a[C_\alpha]$ and $\gammat^\alpha_a = \gamma_a[C^\alpha]$.
Our goal is to compute correlators of the form
\begin{equation}
\la F(\sigma) \gamma^{\alpha_1}_{a_1} \cdots \gamma^{\alpha_k}_{a_k} \gammat^{\beta_1}_{b_1} \cdots \gammat^{\beta_m}_{b_m}\ra.
\end{equation}

As before, we will perform the computations by  localizing the path integral to the vacua and expanding the action to quadratic order in fluctuations.  
Working in a particular vacuum, we can check that the usual decoupling of the non-zero modes holds for the non-local insertions.  In the mode expansion 
given above, we have
\begin{equation}
\int_C \rho = \int_C \sum_{\alpha=1}^h (\omega_\alpha\rho_0^\alpha+\omegat^\alpha\rhot_{0\alpha}) + \sum_k (\ff{1}{\lambda_k} \int_C \ast d f_k) \rhot_k,
\end{equation}
while terms in the action have the schematic form
\begin{equation}
S = \cdots + \lambda_k (\rho_k \chi_k + \rhot_k \theta_k) + i (\bar{m} \theta_k \chi_k + m \rhot_k \rho_k).
\end{equation}
The $\rho_k$ modes do not appear in the observable, since they correspond to exact forms.  This, together with the pairing of the modes in
the action ensures that the terms with $\rhot_k$ will vanish.  Applying the same reasoning to the zero modes shows that non-zero correlators
must have a pairing between insertions of $\gamma_a^\alpha$ and $\gammat_b^\alpha$.   Thus, we can restrict attention to correlators of 
\begin{equation}
\label{eq:pairnonloc}
\Gamma^\alpha_{ab}  = 2 i  \gamma^\alpha_a \gammat^\alpha_b.
\end{equation}
The computation is simplified by noting that the action does not mix modes that correspond to non-intersecting cycles,  so that the contribution
of a particular vacuum will be a product of contributions from the various $\alpha$s.  Fixing to a particular vacuum $\sigma = \sigmah$, and some
choice of $\alpha$, the integral over the $\rho$ zero modes is now a standard finite-dimensional Grassmann integral:
\begin{equation}
\int D[\rhot\rho]_0 \rho_{a_1} \rhot_{b_1} \cdots \rho_{a_k}\rhot_{b_k} e^{-\rhot_{a} \cH^{ab} \rho_b} = 
H \sum_{P\{b_1,\ldots,b_k\}} \ep(P) (\cH^{-1})_{b_{P_1} a_1} \cdots (\cH^{-1})_{b_{P_k} a_k},
\end{equation}
where 
\[ D[\rhot\rho]_0 = \left(\prod_{c=1}^r d\rhot_c d\rho_c\right),\]
$\cH^{ab}$ is the Hessian of the superpotential evaluated at the critical point $\sigmah$, $H = \det \cH$, $P\{b_1,\ldots,b_k\}$ is a permutation
of the set $\{b_1,\ldots,b_k\}$, and $\ep(P)$ is the sign of the permutation.

Using this result for each $\alpha$, and summing over the vacua, we find
\begin{eqnarray}
\label{eq:nonlocans}
\la F(\sigma) \prod_{\alpha\in J} & & \!\!\!\!\!\! \prod_{k=1}^{u^\alpha} \Gamma^\alpha_{a_k b_k} \ra_h =  \sum_{\sigmah} H^{h-1} F \times\nonumber\\
 &\times & \prod_{\alpha\in J} \sum_{P\{b_1\cdots b_{u^\alpha}\}} \ep(P) (\cH^{-1})_{b_{P_1} a_1} \cdots (\cH^{-1})_{ b_{P_{u^\alpha}} a_{u^\alpha}},
\end{eqnarray}
where $J \subseteq \{1,\ldots,h\}$.  

As this general form might be slightly confusing, let us give two useful special cases.  First, consider the case where $\Sigma_h$ is a torus, 
so that there is just a single $\alpha$. The general formula simplifies to
\begin{equation}
\label{eq:nonlocanstorus}
\la F(\sigma) \prod_{k=1}^{u} \Gamma_{a_k b_k} \ra_1 =
\sum_{\sigmah} F \sum_{P\{b_1\cdots b_{u}\}} \ep(P) (\cH^{-1})_{b_{P_1} a_1} \cdots (\cH^{-1})_{ b_{P_u} a_{u}}.
\end{equation}
Second, we can keep the genus of the Riemann surface arbitrary, but take a Landau-Ginzburg theory with just a single multiplet.  Now
the correlators are even simpler:
\begin{equation}
\label{eq:nonlocans1f}
\la F(\sigma) \prod_{\alpha\in J} (\Gamma_\alpha)\ra_{h} = \sum_{\sigmah} H^{h-1-|J|} F.
\end{equation}

We have now completed our goal of computing the non-local observables in the twisted, massive Landau-Ginzburg theory.  So far, this has
been nothing but a simple example of the kinds of structures the reader might wish to study in more sophisticated CTFTs.  In the next section 
we will repay some of the reader's patience by showing that this simple analysis can be carried over with minimal changes to a set of richer
topological theories:  the twisted Gauged Linear Sigma Models for compact toric target-spaces.

\section{Compact Toric Gauged Linear Sigma Models} \label{s:glsm}
The Gauged Linear Sigma Model (GLSM) was introduced by Witten in \cite{W:phases} and has seen many applications in
the last fourteen years.  The GLSM is a two-dimensional $N=(2,2)$ SUSY gauge theory with $n$ chiral multiplets coupled to
$r$ abelian gauge fields.  In addition to the minimal gauge couplings, the model depends upon a choice of a Fayet-Ilioupoulos
parameter $r^a$ and a $\theta$-angle $\theta^a$ for each of the $r$ $\GU[1]$ factors.  They enter the action through holomorphic
couplings $\tau^a = i r^a + \theta^a/2\pi$ in the {\em twisted} superpotential.  The model may be generalized further by introducing
a gauge-invariant superpotential for the matter fields.  We will set the matter superpotential to zero, and, for reasons that
will be clear shortly, we will refer to such GLSMs as {\em toric}.  In the untwisted theory, the $r^a$ are not really 
parameters---they run under the RG flow, leading to quite a bit of interesting physics \cite{APS:tach,Vafa:tachyons,HKMM:tachyons}.  
We will work in the twisted theory, where these may really be thought of as parameters.

\subsection{A Brief Review of  GLSM ``Phases''}
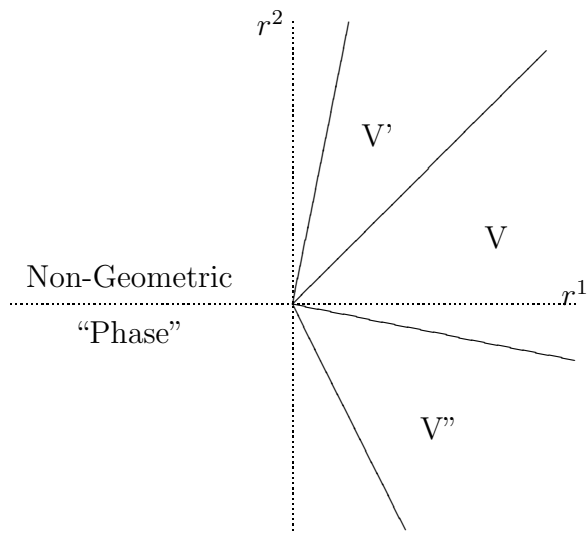
\begin{figure}
\[
\begin{xy} <0.75mm,0mm>:
  (0,0)*{} ="o",    (50,0)*{}="1",     (45,45)*{}="2", 
  (10,50)*{}="3", (20,-40)*{}="4",  (0,50)*{}="5",  
  (0,-40)*{}="6",  (-50,0)*{}="7",    (50,-10)*{}="8",
  (50,2)*{r^1}, (-4,50)*{r^2},
  (15,30)*{\text{V'}},
  (36,12)*{\text{V}},
  (26,-22)*{\text{V''}},
  (-29, 5)*{\text{Non-Geometric}},
  (-29,-5)*{\text{``Phase''}},
\ar@{-}|{} "o"; "8" \ar@{-}|{} "o"; "2"
\ar@{-}|{} "o"; "3" \ar@{-}|{} "o"; "4"
\ar@{.}|{} "5"; "6" \ar@{.}|{} "7"; "1"
\end{xy}
\]
\caption{Phases of the GLSM: a two-parameter example.}
\label{fig:phases}
\end{figure}

Many basic properties of the toric GLSM follow from the structure of the moduli space of classical vacua.  This moduli space 
is obtained by solving the $D$-terms and identifying gauge equivalent points.  There are a number of excellent papers
that describe the resulting structure, for example \cite{W:phases,MP:summing}, so we will be brief here.  The upshot
is that the moduli space depends on the Fayet-Iliopoulos parameters through their appearance in the D-terms.   At a generic point 
in the parameter space, the gauge group is completely broken, and the light degrees of freedom are to be found among the un-eaten 
matter multiplets.  There is a co-dimension one locus where a single $\GU(1)$ becomes un-Higgsed, so that the space $\R^r$ parametrized by 
the $r^a$ is partitioned  into ``phases'', as shown in figure \ref{fig:phases}. 

In general, one finds a number of phases, where the corresponding classical moduli spaces of vacua are birationally equivalent toric 
varieties of complex dimension $n-r$ $V$, $V'$, $V''$, etc.  It is standard to refer to a given phase by the corresponding toric variety.  
One can argue that deep 
in the interior of the cone corresponding to any of these phases, the low energy theory of the GLSM corresponds to a Non-Linear Sigma 
Model (NLSM) with the corresponding toric variety as the target-space .   It is not hard to show that any toric variety with a simplicial
toric fan can be realized as a phase of a GLSM.  

Taking quantum corrections into consideration shows that the ``boundaries'' between 
the phases do not  correspond to real co-dimension one singularities.  Instead, the classical singularities associated to 
the massless gauge multiplets are either smoothed  out by quantum effects, or at worst occur only at particular values of the $\theta$ 
angles\cite{W:phases}.  Thus, the ``phases'' are all smoothly connected, and we expect that the topologically twisted theories obtained from 
NLSMs with target-spaces $V$,$V'$,$V''$ are simply different semi-classical expansions of the same theory.  That is, supposing one can compute
the correlators in the $V$ NLSM, one can analytically continue in the parameters of $V$ to the region where the $V'$ NLSM provides 
a better semi-classical description.  This statement is particularly powerful in the TFT context, where we expect semi-classical
approximations to be exact.

When $V$ is compact, we call the GLSM {\em compact}.  All compact toric GLSMs have an important common feature: the parameter 
space is not covered by the geometric phases.  There is always a ``non-geometric'' phase, where there are no solutions to the classical 
$D$-terms.  The SUSY breaking in the non-geometric phase is merely a classical illusion.  As was already
described in the original work of Witten \cite{W:phases}, in addition to the Higgs vacua described above, the model also has Coulomb
vacua, where the complex scalars in the gauge multiplets acquire non-zero expectation values and give masses to the matter 
fields.\footnote{In non-compact GLSMs, these Coulomb vacua may even be present in what one may have thought of as
a geometric phase\cite{ME:glsmtach,ME:atoric}.  In order to use a semi-classical expansion about the vacua of the GLSM
to compute the $A$-model correlators, one would have to sum over the Higgs vacua (the gauge instantons) {\em and} the Coulomb
vacua.  This can lead to interesting consequences such as the violation of quantum cohomology relations.}  In a compact toric GLSM
the SUSY vacua in the non-geometric phase are massive Coulomb vacua.  These vacua can be given an effective description by integrating
out the massive matter fields.   This yields the famous effective twisted superpotential 
\begin{equation}
\label{eq:superpot}
W = \sum_{a=1}^r \sigma_a \log \left[ \prod_{i=1}^n \left(\frac{1}{\exp(1)\mu} \sum_{b=1}^r Q_i^b \sigma_b \right)^{Q_i^a} /q_a\right], 
~~~ q_a = e^{2\pi i \tau^a},
\end{equation}
where $Q_i^a$ are the charges of the $n$ matter fields under the $[\GU(1)]^r$ gauge group, $\sigma_a$ are the complex 
scalars in the gauge multiplets, and $\mu$ is a renormalization scale.  The renormalization scale does not play a role in the 
topological theory and will be set to one in what follows.  The compactness of $V$ ensures that the $\sigma$-vacua obtained from $W$
are massive.

\subsection{Non-Local Observables in the Toric GLSM}
We now come to a simple point: the non-geometric phase with its isolated Coulomb vacua provides another semi-classical description
of the $A$-twisted theory.  This description is by far the simplest one for the purpose of computing the topological correlators.   The 
semi-classical computation  reduces to the massive Landau-Ginzburg theory we studied above, with the same fields, action of $Q$, and 
observables, but with the particular  superpotential of eqn. (\ref{eq:superpot}) and a change in the path integral measure  
generated by integrating out the zero modes of the {\em matter} fields.  This change in the measure was computed in \cite{ME:atoric},
leading to the general formula for the correlators of the local observables:
\begin{equation}
\label{eq:corrsGLSM}
 \la \sigma_{a_1}(x_1) \cdots \sigma_{a_k}(x_k)\ra_h  =  \sum_{\sigmah} [\prod_{i=1}^n \xi_i]^{h-1} [H]^{h-1}
\sigmah_{a_1} \cdots \sigmah_{a_k},
\end{equation}
where $\xi_i = \sum_{a=1}^r Q_i^a \sigma_a$, and $H= \det\Hess W$ as before.

In fact, nothing in \cite{ME:atoric} assumed that we were computing correlators of local observables, and the argument can be repeated 
verbatim for correlators involving one-form observables.  Since we have learned how to compute  correlators of non-local observables in 
the Landau-Ginzburg theory described below, we now know how to compute these in the GLSM:  one should use the superpotential corresponding
to the particular GLSM, and one should insert the measure factor of \cite{ME:atoric}.    Thus, for any compact toric GLSM, we have
\begin{eqnarray}
\label{eq:allGLSMcorrs}
\la F(\sigma) \prod_{\alpha\in J} & & \!\!\!\!\!\! \prod_{k=1}^{u^\alpha} \Gamma_\alpha^{a_k b_k} \ra_h =  \sum_{\sigmah|d W(\sigmah) = 0} [\prod_i \xi_i H]^{h-1} F \times\nonumber\\
 &\times & \prod_{\alpha\in J} \sum_{P\{b_1\cdots b_{u^\alpha}\}} \ep(P) (\cH^{-1})_{b_{P_1} a_1} \cdots (\cH^{-1})_{ b_{P_{u^\alpha}} a_{u^\alpha}}.
\end{eqnarray}
If the reader is eager to apply this formula to specific examples, she might find it helpful to note that the Hessian of the 
superpotential in eqn. (\ref{eq:superpot}) has a simple form:
\begin{equation}
\cH^{ab} = \sum_i \frac{Q_i^a Q_i^b}{\xi_i}.
\end{equation}

\section{Some Examples} \label{s:exmpl}
We will now apply our general results to two simple theories.  The first is a plain Landau-Ginzburg model, while the second is the simplest
compact toric GLSM.  As we will see, these TFTs have identical chiral rings and local correlators, but they 
differ in correlators of the non-local observables.

\subsection{A One Field Landau-Ginzburg Model}
We consider a model with a single multiplet and superpotential depending on a single parameter $q$:
\begin{equation}
\label{eq:LGW}
W = \ff{1}{n+1} \sigma^{n+1} - q \sigma.
\end{equation}
Clearly the model has isolated vacua $\sigmah$ satisfying $\sigmah^n = q$ and a chiral ring of observables $\sigma^a$, with
$0\le a \le n-1$.  From eqn. (\ref{eq:loccorrs}) we have the formula for correlators of local observables:
\begin{equation}
\label{eq:LGlocal}
\la F(\sigma) \ra_h^{\text{LG}} = \sum_{\sigmah} \left( n \sigmah^{n-1}\right)^{h-1} F(\sigmah).
\end{equation}
In particular, given two observables $\sigma^a$, $\sigma^b$, $0\le a,b\le n-1$ we can define the ``TFT metric'' \cite{W:tphase} 
via the two-point function on the sphere:
\begin{equation}
\label{eq:LGmetric}
G_{ab} = \la \sigma^a\sigma^b\ra_0^{\text{LG}} = \delta^{a+b,n-1}.
\end{equation}

Now let us compute some simple correlators of non-local observables: 
\begin{equation}
\la \prod_{\alpha \in J} \Gamma^\alpha\ra^{\text{LG}}_h = \sum_{\sigmah} (n \sigmah^{n-1})^{h-1-|J|}.
\end{equation}
Clearly, non-zero correlators must have $(n-1)(h-1-|J|) = 0 \mod n$, which is only possible if
$h-1-|J| = n m$ for some $m$.  If this holds, we have
\begin{equation}
\la \prod_{\alpha \in J} \Gamma^\alpha\ra^{\text{LG}}_h = n^{nm+1} q^{(n-1)m}.
\end{equation}
On the torus the non-vanishing correlators are :
\begin{equation}
\la \sigma^s \Gamma \ra_1^{\text{LG}} = \delta^{s,n-1}~~~\text{for}~~ 0 \le s \le n-1.
\end{equation}
These will be useful for our discussion of factorization properties of correlators.

We will now compare the observables in this theory with those of a compact toric GLSM: the $\C\P^{n-1}$ model.

\subsection{The $\C\P^{n-1}$ GLSM}
This is the simplest compact toric GLSM.  It is described by $n$ matter multiplets coupled to a single gauge multiplet
with charges $Q_i = 1$, $i=1,\ldots, n$.  Plugging this into the effective superpotential of eqn. (\ref{eq:superpot}), we find
that the $\sigma$-vacua are described by $\sigmah^n = q$, while the measure factors are given by
\begin{equation}
\prod_i \xi_i  = \sigma^n, ~~\text{and}~~ H = \frac{n}{\sigma}.
\end{equation}
Thus, the model has the same chiral ring as the Landau-Ginzburg theory above, and, in fact, identical correlators of 
local observables:
\begin{equation}
\label{eq:GLSMlocal}
\la F(\sigma) \ra_h^{\text{GLSM}} = \sum_\sigmah \left( \sigmah^n \cdot \frac{n}{\sigmah}\right)^{h-1} F(\sigmah) = \la F(\sigma) \ra_h^{\text{LG}}.
\end{equation}
Obviously, the TFT metric is also the same: $G_{ab}^{\text{GLSM}} = G_{ab}^{\text{LG}}$.

Now let us compute correlators of non-local observables by using eqn. (\ref{eq:allGLSMcorrs}).  We find
\begin{equation}
\la \prod_{\alpha \in J} \Gamma^\alpha\ra_{h}^{\text{GLSM}} = \sum_\sigmah \left(\sigmah^n\right)^{h-1} \left(\ff{n}{\sigmah}\right)^{h-1-|J|}
 = q^{h-1} \sum_\sigmah \left(\ff{n}{\sigmah}\right)^{h-1-|J|}.
\end{equation}
Of course, non-zero correlators satisfy $h-1-|J| = n m$, and these are given by
\begin{equation}
\label{eq:GLSMnonloc}
\la \prod_{\alpha \in J} \Gamma^\alpha\ra_{h}^{\text{GLSM}} = n^{nm +1} q^{h-1-m}.
\end{equation}

For completeness, we also give the correlators on the torus:
\begin{equation}
\la \sigma^s \Gamma \ra_{1}^{\text{GLSM}} = q \delta^{s,n-1}, ~~ \text{for}~0\le s\le n-1.
\end{equation}

Comparing these two simple examples, we see that, as promised, we have two TFTs with identical chiral rings and 
correlators of local observables, which nevertheless differ in more general correlators.  We have before us another example of 
the adage \cite{AM:chiral}  ``Chiral rings do not suffice.''

\subsection{Ghost Number Selection Rules}
The non-equivalence of these two models could have been guessed from the selection rules imposed by the anomalous ghost 
number symmetry of the $A$-model.  We will now discuss the selection rules for the two examples and verify that our explicit
computations are consistent with these.

\subsubsection{Ghost Number in the Landau-Ginzburg Theory }
The action of the twisted Landau-Ginzburg theory is invariant under a $\GU(1)$ symmetry with charges
%
\begin{eqnarray}
\sigma & \to & e^{i\alpha} \sigma, \nonumber\\
q          &\to & e^{in\alpha} q, \nonumber\\
\rho      & \to & e^{-i(n-1)\alpha/2}\rho, \nonumber \\
\chi      &\to & e^{+i(n-1)\alpha/2}\chi, \nonumber \\
\theta   &\to & e^{+i(n-1)\alpha/2} \theta.
\end{eqnarray}
It is easy to see that this is  consistent with the action of $Q$ if we assign it charge $(n+1)/2$.
Upon performing this change of variables in the path integral, one finds that the measure picks up an overall factor 
of $e^{i\alpha(1-h)(n-1)}$, which is nothing other than the familiar gravitational anomaly term.  The anomaly plays no
role for $h=1$, and we immediately obtain the selection rule
\begin{equation}
\la \sigma^s \Gamma\ra_1^{\text{LG}} (q) = e^{i\alpha(s-n+1)} \la \sigma^s \Gamma\ra_1^{\text{LG}} (q e^{-in\alpha}).
\end{equation}
Since the correlator is independent of $\bar{q}$, it follows from the selection rule that it must be proportional to $q^A$, $A\in \Z$ 
satisfying $s-n+1 -nA = 0$.  For $0\le s \le n-1$ the only non-zero correlators must have $s=n-1$ and $A=0$, as we found above.

\subsubsection{Ghost Number in the GLSM}
The twisted GLSM has a classical ghost number symmetry with\cite{W:phases,MP:summing}
\begin{eqnarray}
\sigma &\to& e^{i\alpha} \sigma, \nonumber\\
\rho &\to& e^{i\alpha/2} \rho, \nonumber \\
\chi &\to& e^{-i\alpha/2}\chi, \nonumber \\
\theta &\to& e^{-i\alpha/2} \theta, \nonumber\\
Q  & \to & e^{i\alpha/2} Q.
\end{eqnarray}
This classical symmetry is violated by two quantum effects: the gauge anomaly and the gravitational anomaly.
The effect of the former can be absorbed into a shift of the $\theta$ angle, leading to $q \to e^{in\alpha} q$, while
the latter simply gives an over-all factor of $e^{-i\alpha(1-h)(n-1)}$ in the transformation of the measure.  All together,
we find the following selection rule for local correlators:
\begin{equation}
\la \sigma^s \ra_h^{\text{GLSM}}(q) = e^{i\alpha s} e^{-i\alpha(1-h)(n-1)} \la \sigma^s \ra_h^{\text{GLSM}} (q e^{-i\alpha n}).
\end{equation}
Using holomoprhy in $q$, it is easy to see that the correlator is proportional to $q^A$, and the integer $A$ 
must satisfy $s=(1-h)(n-1)+nA$.  This is a selection rule familiar in the $\C\P^{n-1}$ model. 

The selection rule for correlators with a non-local insertion is  
\begin{equation}
\la \sigma^s \Gamma\ra_1^{\text{GLSM}} (q) = e^{i\alpha(s+1)} \la \sigma^s \Gamma\ra_1^{\text{GLSM}} (q e^{-in\alpha}).
\end{equation}
Holomorphy in $q$ again implies the $q^A$ form, and for $0\le s \le n-1$, the only non-zero correlator that is allowed 
must have $s=n-1$ and $A=1$, which is what we found in our explicit analysis.

It is instructive to compare the selection rule from the gauge theory perspective (i.e. a semi-classical expansion in
the geometric phase) to the ``L-G'' point of view (i.e. the semi-classical expansion in the non-geometric phase).
The anomalous breaking of the symmetry is now replaced by explicit breaking via $W$, and invariance of the
action can be restored by assigning charge $n$ to $q$.  Remembering the additional transformation of the
measure due to the $[\prod_i \xi_i]^{h-1}$ factor, it is easy to reproduce the GLSM selection rules above.

Comparing the charges of $\sigma$ and $Q$ in the Landau-Ginzburg theory to those in the $\C\P^{n-1}$ GLSM, we
see that the descendants of $\sigma$ have different ghost numbers in the two theories, and there is no reason for
their correlators to agree, and in fact they should disagree in precisely the manner we found by explicit computation.
The explicit computation simply verifies the (entirely pedantic) point that the coefficients of $q^A$ are non-zero.

\section{Discussion} \label{s:discus}

\subsection{Mathematical Properties of the Correlators}
We have presented a study of some non-local observables in a large class of $A$-model TFTs.  We hope we have 
convinced the reader that this class of models presents a tractable setting in which to investigate such observables.
The most useful result we have obtained is the expression for correlators in any compact toric GLSM.  It is sufficiently
simple that it should be easy apply in models with two, and maybe even three parameters to give closed-form expressions for
the correlators.  However, even having the form as the sum over the roots of the polynomial system has a number of
useful consequences.  For example, it is not hard to argue that the correlators are rational functions of the parameters $q_a$,
and it is obvious that the quantum cohomology relations of the GLSM hold.  One suspects that further vanishing theorems
for insertions of non-local observables could be found.  Furthermore, it should be possible to recast these more general
correlators as some (toric) residue, much as can be done for the Landau-Ginzburg theories \cite{Vafa:tlg} or for the 
GLSM\footnote{For local observables this can be seen by comparing the form of the local correlators in eqn. (\ref{eq:corrsGLSM}) with the Horn
uniformization formula of GKZ\cite{GKZ:horn} and the toric residue formulas found in \cite{BM:torres,KARU:torres,SV:torres}.  We thank 
E. Materov, K. Karu and M. Vergne for discussions on this point.}.  We leave these questions for future investigations.

\subsection{Factorization Properties of the Correlators}
An important motivation for this work was a frustration with a wonderful property of correlators of local observables in TFT: 
factorization \cite{W:tphase}.  This property reduces local correlators on $\Sigma_h$ to the TFT metric and three-point functions on the sphere.
Thus, in some sense, the computation for $h >0$ is vacuous.  As we will now argue, even with non-local insertions, computations 
with $h >1$ are still vacuous.  However, we have at least decreased our frustration by an integral amount.

There are two relations that allow one to reduce computations of local observables on a Riemann surface to 
computations on surfaces of lower genus.  In the first, a $\Sigma_h$
is split into a $\Sigma_{h'}$ and $\Sigma_{h-h'}$.  Supposing that $\{\O_i\}$ are a basis for the local observables, and we can write
$F(\sigma) = f(\sigma) g(\sigma)$, it has been shown that
\begin{equation}
\label{eq:factor1}
\la F(\sigma) \ra_h = \sum_{ij} \la f(\sigma) \O_i \ra_{h'} G^{ij} \la \O_j g(\sigma) \ra_{h-h'},
\end{equation}
where $G^{ij}$ is the inverse of the TFT metric $G_{ij} = \la \O_i \O_j\ra_{0}$.
The second relation allows us to pinch a cycle in $\Sigma_h$ to obtain $\Sigma_{h-1}$:
\begin{equation}
\label{eq:factor2}
\la F(\sigma) \ra_h = \sum_{ij} \la F(\sigma)\O_i \O_j \ra_{h-1} G^{ij}
\end{equation}

An insertion of a non-local observable will  invalidate the second relation, since the number of distinct non-local observables
on $\Sigma_h$ is proportional to $h$.  However, we still expect the first relation to hold.  After all, we can
choose a metric on $\Sigma_h$ so that a long thin tube separates the $\Sigma_{h'}$ and $\Sigma_{h-h'}$ components, and
we can choose representatives for observables, local, as well as non-local, that are well separated from this tube.  As we 
make the tube longer and longer, the non-local insertions stay well separated, and we expect exactly the same reasoning as 
for local operators to yield eqn. (\ref{eq:factor1}).  

It is clear that we can use the remaining factorization property to reduce the correlators to computations on the torus. It is simple
and instructive to check that this property indeed holds for the two examples we considered above.  In each of these, the chiral
ring of local observables is given by $\sigma^i$, $0\le i \le n-1$, and applying the factorization rule we expect
\begin{equation}
\la \Gamma_1 \cdots \Gamma_k \ra_h =  \la \Gamma \sigma^{i} \ra_1 G^{i j} \la\sigma^{j} \Gamma_2 \cdots \Gamma_k \ra_{h-1}
\end{equation}
Using the explicit forms for  the metric and the correlators at genus one, it is easy to see that the property does indeed hold.  

\subsection{Two-Form Observables}
The reader may wonder whether the simple computations for the one-form observables readily extend to the two-form observables.
While there is no problem in principle of applying the localization techniques to correlators with such insertions, there are a number
of technical problems associated to the use of equations of motion and the appearance of ``interactions'' in the two-form observables
themselves.  We saw this explicitly in the Landau-Ginzburg theory, where $\O_{f(2)}$ explicitly involved the superpotential.  As the
computation in the Appendix illustrates, one can still compute correlators of such operators via localization, but the computation is
more involved, and one will certainly not be able to provide as clean an answer as for correlators with $\O_{f(1)}$ insertions.  

We expect the same issues to arise in the GLSM, where instead of the superpotential we will find matter fields in $\O_{f(2)}$.  These
terms will have an interesting consequence: unlike for correlators of local and one-form observables, as soon as there are 
insertions of the $\int_{\Sigma_h} \O_{f(2)}$, we will not be able to simply absorb the matter zero modes into an over-all measure factor in an
effective Landau-Ginzburg computation.  To compute correlators of these objects, one will have to repeat the analysis of
\cite{ME:atoric} and carefully treat the matter zero modes both in the measure and in the insertions of the two-form observables. 

\subsection{In Search of Geometric Meaning}
We would have liked to make a clear connection between these correlators and some invariants of the corresponding manifolds.
Unfortunately, it is not entirely clear how to do this, since our discussion is restricted to TFT and does not discuss coupling the theory
to two-dimensional gravity.  We are currently studying the proper framework for this coupling, and we suspect that these results will
find a proper geometric interpretation once gravity is properly taken into account.

\acknowledgments
I would like to thank M.R. Plesser for a number of useful discussions and suggestions for this project, and I am grateful to N. Halmagyi for useful
comments on the manuscript.  This article is based upon work supported 
in part by the National Science Foundation under Grants PHY-0094328 and PHY-0401814 .  Any opinions, findings, and conclusions or 
recommendations expressed in this article are those of the author and do not necessarily reflect the views of the National Science 
Foundation.

\appendix
\section{Two-Form Observables in the Landau-Ginzburg TFT} \label{s:app1}
It is fairly easy to 
generalize the localization techniques used above to compute correlators with insertions of two-form observables.  This is
particularly straightforward for the Landau-Ginzburg theory, while for the GLSM it would involve re-visiting the matter zero
modes.  We will leave the latter case for future work and here merely describe the simpler case of Landau-Ginzburg TFTs.

From our general discussion of descent in CTFT, we expect that $\O_{f(2)}$ may be used to deform the topological field
theory.  In fact, it is easy to see that in the Landau-Ginzburg case, the corresponding deformation is just a change in the
superpotential:  $W\to W+f$.  We can see this by computing the change in the action under a change 
in the superpotential:
\begin{eqnarray}
\label{eq:dSdw}
-\delta_W S & = & \int_{\Sigma_h} \{ i \rho_a\wedge\rho_b \delta W_{,ab} - \delta W_{,a} \ast \Wb_{,a} \}\nonumber\\
~ & ~ & - \int_{\Sigma_h} \ast \{ \delta \Wb_{,a} W_{,a} + 2 \chi_a \delta \Wb_{,ab} \theta_b \}.
\end{eqnarray}
The first line is recognized as $\O_{\delta W (2)}$, while the second is $Q$-exact, so that
\begin{equation}
\label{eq:dSdW2}
-\delta_W S =  \O_{\delta W(2)} + \{Q, \int_{\Sigma_h} \ast \chi_a \delta \Wb_{,a} \}.
\end{equation}

Since we have the explicit form of the correlators for local observables in eqn. (\ref{eq:loccorrs}) we can carry out an amusing and
instructive exercise of comparing the first order deformation of the superpotential in the explicit formula to the 
correlator with an additional insertion of $\int_{\Sigma_h} \O_{\delta W(2)}$.  This computation will also demonstrate how localization may be
used to compute correlators with arbitrary insertions of two-form observables. For simplicity, we will restrict to a one-multiplet 
Landau-Ginzburg theory.  The generalization to several multiplets is straightforward.  

We expect 
\begin{equation}
\label{eq:comparedef}
\la F(\sigma) \O_{\delta W(2)} \ra_h  =  \sum_{\sigmah+\delta\sigmah} [\det\Hess (W+\delta W)]^{h-1} F(\sigma+\delta\sigmah) 
+ O(\delta W^2).
\end{equation}
This indeed holds, but with the important caveat that correlators behave smoothly as one changes $W$ only as long as $\delta W$ 
does not change the large field behavior of the superpotential\cite{Vafa:tlg}.  This is not surprising from the perspective of the 
untwisted Landau-Ginzburg theory:  a change in the large field behavior of $W$ will, in general, cause a jump in the Witten index 
of the theory.  Thus, we should restrict our analysis to $\delta W$ that leaves the large field behavior fixed.  In that case,  no new roots $\sigmah$ 
are produced, and the solutions to $W' = 0$ are simply shifted by
\begin{equation*}
\delta \sigmah = - \delta W'(\sigmah) / W''(\sigmah) + O(\delta W^2).
\end{equation*}
Plugging this into the right-hand side of eqn. (\ref{eq:comparedef}) and expanding to first order in $\delta W$, we find
\begin{eqnarray*}
\delta_W \sum_{\sigmah} \left[ W''(\sigmah)\right]^{h-1} F(\sigmah)  & = &
       (h-1) \la \delta W''(\sigma) F(\sigma)\ra_{h-1} \nonumber\\
 ~ & ~&        
      - \la F'(\sigma) \delta W'(\sigma)\ra_{h-1} 
      -(h-1) \la \delta W'(\sigma) W'''(\sigma) F(\sigma)\ra_{h-2}.
\end{eqnarray*}

Now let us see what can be said about the left-hand side of eqn. (\ref{eq:comparedef}).  It is sufficient to examine the contribution from a particular vacuum $\sigma_v$. 
We will assume that we can reduce the analysis to the zero modes, and we will compute in the $V_g(\Sigma_h) \to \infty$ limit.  We wish to 
compute the contribution of the $\sigma_v$ vacuum to
\begin{equation}
\la F(\sigma) \int_{\Sigma_h} \left[ i \rho \wedge \rho \delta W'' - \ast \delta W' \Wb' \right] \ra_{h;\sigma_v} = \int D[\text{fields}]_0 F(\sigma) e^{-S_0} (\Delta_F + \Delta_B),
\end{equation}
where to $O(1/\sqrt{V_g})$, we have
\begin{eqnarray}
\Delta_F  & = & \int_{\Sigma_h} i \rho \wedge \rho \delta W'' = 
-2 i \delta W'' (\sigma_v) \sum_{\alpha=1}^{h} \rhot_{0\alpha} \rho_0^\alpha, \nonumber\\
\Delta_B & = & -\int_{\Sigma_h} \ast \delta W' \Wb' \nonumber \\ 
&= &~-\sqrt{V_g}~ \delta W'(\sigma_v) \Wb''(\sigma_v) \sigmab_0 
                                                                                   -\delta W''(\sigma_v) \Wb''(\sigma_v) |\sigma_0|^2.
\end{eqnarray}
The most interesting term  in this expansion is the $O(\sqrt{V_g})$ term in $\Delta_B$:  its presence means that terms of $O(1/\sqrt{V_g})$
in $S_0$ and $F(\sigma)$ will contribute to the correlator.  Expanding these to requisite order, one finds
\begin{eqnarray*}
e^{-S_0} & = & e^{-S_0}|_{\sigma_v}  - \frac{e^{-S_0}|_{\sigma_v}}{2\sqrt{V_g}}  \left[ 
 W''' \sigma_0 ( \Wb'' |\sigma_0|^2 -2i \sum_{\alpha=1}^h \rhot_{0\alpha} \rho_0^\alpha)  \right.   \\
  &~&~~~~~~~~~~~~~~\qquad\qquad \left.+  \Wb''' \sigmab_0( W''|\sigma_0|^2  - 2 \chi_0 \theta_0)  \right],
\end{eqnarray*}
and 
\begin{equation*}
F(\sigma) = F(\sigma_v) + F'(\sigma_v) \sigma_0 / \sqrt{V_g}.
\end{equation*}
Finally, plugging these in and carrying out the Gaussian integrals, one finds that the $O(\sqrt{V_g})$ terms vanish, while
the $O(1)$ terms give contributions: the $\Delta_F$ insertion yields $(h-1)\la \delta W'' F(\sigma)\ra_{h-1}$, and 
the $\Delta_B$ insertion yields 
\begin{equation*}
- \la \delta W' F' \ra_{h-1} + (1-h) \la \delta W' W''' F \ra_{h-2}.
\end{equation*}
Putting these together reproduces the expansion of the explicit formula for the correlator.

It is fairly clear that by generalizing this expansion in $\sqrt{V_g}$ one will be able to obtain correlators with any number of
two-form observable insertions.  Of course, the computation will be more involved than for the one-form observables.


\end{document}